# Eigenvalues and Eigenfunctions of Woods‒Saxon Potential in *PT* Symmetric Quantum Mechanics

Ayşe Berkdemir [a], Cüneyt Berkdemir [a] and Ramazan Sever [b*]

[a] *Department of Physics, Faculty of Arts and Sciences, Erciyes University, Kayseri, Turkey*
[b] *Department of Physics, Middle East Technical University, Ankara, Turkey*

**Abstract**

Using the Nikiforov Uvarov method, we obtained the eigenvalues and eigenfunctions of the Woods‒Saxon potential with the negative energy levels based on the mathematical approach. According to the *PT*‒ Symmetric quantum mechanics, we exactly solved the time‒independent Shcrödinger equation for the same potential. Results are obtained for the s-states.

[*]Corresponding author: Ramazan Sever, E-mail: sever@metu.edu.tr

## 1. Introduction

In the appearance of non-relativistic quantum mechanics, one usually chooses a real

(Hermitian) potential to derive the real energy eigenvalues of the corresponding time – independent Schrödinger equation [1]. About six years ago, Bender and his co-workers and later others have studied several complex potentials on the *PT* - symmetric quantum mechanics. They showed that the energy eigenvalues of the Schrödinger equation are real [2,3,4,5,6,7,8]. Afterwards non – Hermitian Hamiltonians with real or complex spectra have been studied by using numerical and analytical techniques [9,10,11].

Various different methods have been adopted for the solution of the above mentioned potential cases. Such as, one of these methods which makes it possible to present the theory of special functions by starting from a differential equation has been developed by Nikiforov and Uvarov Method(NU) [12]. This method is based on solving the time – independent Schrödinger equation by reduction to a generalized equation of hypergeometric type.

In this work, the Schrödinger equation is solved by using the NU method to ensure the energy eigenvalues and eigenfunctions of bound state for real and complex form of the Woods – Saxon potential. It is selected for a shell model for describing metallic clusters in a successful way and for lighting the central part of the interaction neutron with one heavy nucleus [13,14].

This paper is arranged as follows: In Sec. II we briefly review Nikiforov-Uvarov method. In Sec. III solution of the Schrödinger equation with Woods-Saxon potential is introduced first and later with PT-symmetric and non-PT-symmetric non-Hermitian forms of the q-deformed forms are obtained. In Sec. IV we discuss the results.

## 2. Nikiforov-Uvarov Method

NU method is based on the solution of a general second order linear differential equation with special orthogonal functions. So a non-relativistic Schrödinger equation can be solved by using this method. Thus, for a given real or complex potential, the Schrödinger equation in one dimension is reduced to a generalized equation of hypergeometric type with an appropriate $s = s(x)$ coordinate transformation. Therefore it can be written in the following form,

$$\psi''(s) + \frac{\tilde{\tau}(s)}{\sigma}\psi'(s) + \frac{\tilde{\sigma}(s)}{\sigma^2(s)}\psi(s) = 0 \qquad (1)$$

where $\sigma(s)$ and $\tilde{\sigma}(s)$ are polynomials, at most second-degree, and $\tilde{\tau}(s)$ is a first-degree polynomial. Hence Schrödinger equation and the Schrödinger like equations can be solved by means of the special potentials or some quantum mechanical problems.

By using the transformation

$$\psi(s) = \phi(s)y(s) \qquad (2)$$

Eq.(1) reduces into a hypergeometric type form

$$\sigma(s)y'' + \tau(s)y' + \lambda y = 0 \qquad (3)$$

where $\phi(s)$ satisfies $\phi(s)'/\phi(s) = \pi(s)/\sigma(s)$. $y(s)$ is the hypergeometric type function whose polynomial solutions are given by Rodrigues relation

$$y_n(s) = \frac{B_n}{\rho(s)}\frac{d^n}{ds^n}\left[\sigma^n(s)\rho(s)\right], \qquad (4)$$

where $B_n$ is a normalizing constant and the weight function $\rho$ must satisfy the condition in Ref.[12]

$$(\sigma\rho)' = \tau\rho. \qquad (5)$$

The function $\pi$ and the parameter $\lambda$ required for this method are defined as

$$\pi = \frac{\sigma' - \tilde{\tau}}{2} \pm \sqrt{\left(\frac{\sigma' - \tilde{\tau}}{2}\right)^2 - \tilde{\sigma} + k\sigma} \qquad (6)$$

and
$$\lambda = k + \pi'. \qquad (7)$$

On the other hand, in order to find the value of $k$, the expression under the square root must be square of polynomial. Thus, a new eigenvalue equation for the Schrödinger equation becomes

$$\lambda = \lambda_n = -n\tau' - \frac{n(n-1)}{2}\sigma'', \quad (n = 0,1,2,...) \qquad (8)$$

where
$$\tau(s) = \tilde{\tau}(s) + 2\pi(s), \qquad (9)$$

and it must have a negative derivative.

## 3. Woods–Saxon Potential

A basic problem in the nuclear physics is the motion of the free electrons which have a conclusive influence on the abundance of metallic clusters. These electrons are moving in well defined orbitals, around the central nucleus and in a mean field potential which is produced by the positively charged ions and the rest of the electrons. In the mean field potential, the details of the potential are described by free parameters such as depth, width and the slope of the potential, which have to be fitted to experimental observation. Therefore, a mean field potential is always empirical and its an example can be given as the Woods–Saxon potential [15]

$$V(r) = \frac{-V_0}{1 + e^{\left(\frac{r-R_0}{a}\right)}}, \qquad (10)$$

where $V_0$ is the potential depth, $R_0$ is the width of the potential and a its diffuseness and $a$ is the surface thickness which is usually adjusted to the experimental values of ionization energies.

Schrödinger equation in the spherical coordinates is

$$\left(-\frac{\hbar^2}{2m}\nabla^2 + V(r)\right)\psi(r) = E\psi(r) \qquad (11)$$

or the radial part of it takes the form

$$R''(r) + \frac{2m}{\hbar^2}\left[E + \frac{V_0}{1 + qe^{2\alpha r}}\right]R(r) = 0 \qquad (12)$$

where the radial wave function $\psi(r)$ is written as $\psi(r) = R(r)/r$ and the conversions of $r - R_0 \equiv r$, $1/a \equiv 2\alpha$ are done by inserting an arbitrary real constant $q$ within the potential. In addition, here we assume that $\psi(r) = (1/r)R(r)$ is bounded as $r \to 0$.

Now, we apply a transformation to $s = -e^{2\alpha r}$ to get a form that NU method applicable. Thus by introducing the following dimensional parameters

$$\varepsilon = -\frac{mE}{2\hbar^2\alpha^2} > 0 \quad (E < 0), \qquad \beta = \frac{mV_0}{2\hbar^2\alpha^2} \quad (\beta > 0), \qquad (13)$$

which leads to the generalized of hypergeometric type given by Eq.(1):

$$\frac{d^2 R(s)}{ds^2} + \frac{1-qs}{s(1-qs)}\frac{dR(s)}{ds} + \frac{1}{s^2(1-qs)^2} \times \left[-\varepsilon q^2 s^2 + (2\varepsilon q - \beta q)s + \beta - \varepsilon\right]R(s) = 0. \qquad (14)$$

After the comparison of Eq.(14) with Eq.(1), we obtain the corresponding polynomials as
$$\tilde{\tau}(s) = 1 - qs, \quad \sigma(s) = s(1-qs), \quad \tilde{\sigma}(s) = -\varepsilon q^2 s^2 + (2\varepsilon q - \beta q)s + \beta - \varepsilon. \tag{15}$$
Substituting these polynomials in Eq.(6), we achieve $\pi$ function as
$$\pi(s) = -\frac{qs}{2} \pm \frac{1}{2}\sqrt{(q^2 + 4\varepsilon q^2 - 4kq)s^2 + 4(\beta q - 2\varepsilon q + k)s + 4(\varepsilon - \beta)} \tag{16}$$
taking $\sigma'(s) = 1 - 2qs$. The constant $k$ is determined from the corresponding statement in Ref. [12], i.e., $k_{1,2} = \beta q \pm q\sqrt{(\varepsilon - \beta)}$. Afterwards, these two values for each $k$ are replaced in Eq.(16) and the following possible solutions for $\pi(s)$ are obtained as
$$\pi(s) = -\frac{qs}{2} \pm \frac{1}{2}\begin{cases} (2\sqrt{\varepsilon-\beta}-1)qs - 2\sqrt{\varepsilon-\beta} & \text{for} \quad k=\beta q + q\sqrt{\varepsilon-\beta}, \\ (2\sqrt{\varepsilon-\beta}+1)qs - 2\sqrt{\varepsilon-\beta} & \text{for} \quad k=\beta q - q\sqrt{\varepsilon-\beta} \end{cases} \tag{17}$$
It is clearly seen that the energy eigenvalues are found with a comparison of Eq.(7) and Eq.(8). Therefore, the polynomial $\tau(s)$ in Eq.(9), for which its derivative has a negative value, is established by a suitable choice of the polynomial $\pi(s)$ for $k = \beta q - q\sqrt{\varepsilon - \beta}$ from Eq.(17):
$$\pi(s) = -\frac{qs}{2} - \frac{1}{2}\left[(2\sqrt{\varepsilon-\beta}+1)qs - 2\sqrt{\varepsilon-\beta}\right], \tag{18}$$

$$\tau(s) = 1 + 2\sqrt{\varepsilon-\beta} - q(3 + 2\sqrt{\varepsilon-\beta})s, \quad \tau'(s) = -q(3 + 2\sqrt{\varepsilon-\beta}). \tag{19}$$
After substituting the polynomials $\pi(s)$ and $\tau'(s)$ and also $k$ we get
$$\lambda = q(\beta - 2\sqrt{\varepsilon-\beta} - 1), \tag{20}$$
and
$$\lambda = \lambda_n = nq(3 + 2\sqrt{\varepsilon-\beta}) + n(n-1)q, \tag{21}$$
Thus the parameter $\varepsilon$ takes the form
$$\varepsilon_n = \left(\frac{\beta}{2(n+1)}\right)^2 + \left(\frac{n+1}{2}\right)^2 + \frac{\beta}{2}. \tag{22}$$
Substituting the values of $\varepsilon$ and $\beta$ from Eq.(13) and the transformation $2\alpha \equiv 1/a$ in Eq(22), one can immediately determine the energy eigenvalues $E_n$ as
$$E_n = -\frac{\hbar^2}{2ma^2}\left[\left(\frac{ma^2 V_0}{\hbar^2(n+1)}\right)^2 + \left(\frac{n+1}{2}\right)^2 + \frac{ma^2 V_0}{\hbar^2}\right]. \tag{23}$$
Here, the index $n$ is non-negative integers with $\infty > n \geq 0$ and the above equation indicates that we deal with a family of the standard Woods–Saxon potential. Of course it is clear that by imposing appropriate changes in the parameters $a$ and $V_0$, the index $n$ describes the quantization of the bound states and the energy spectrum. It is illustrated in Fig.1, the Woods–Saxon potential given by Eq.(10) and some of the initial energy levels for different $q$ values are presented by choosing $V_0 = 5\ MeV$, $R_0 = 5.8\ fm$ and $a = 0.65\ fm$.

Fig.2 shows that the energy eigenvalues as a function the discrete level $n$ for different values of the parameter $a$.

We now find the corresponding eigenfunctions. The polynomial solution of the hypergeometric function $y(s)$ depend on the determination of weight function $\rho(s)$. Thus $\rho(s)$ in Eq.(5) is calculated as

$$\rho(s) = (1-qs)s^{2\sqrt{\varepsilon-\beta}}, \qquad (24)$$

and substituting in the Rodrigues relation in Eq.(4), the polynomial $y(s)$ is organized in the following form

$$y_n(s) = B_n(1-qs)^{-1}s^{-2\sqrt{\varepsilon-\beta}}\frac{d^n}{ds^n}\left[(1-qs)^{n+1}s^{n+2\sqrt{\varepsilon-\beta}}\right]. \qquad (25)$$

Choosing $q=1$, the polynomial solution of $y_n(s)$ is expressed in terms of Jacobi Polynomials, which is one of the orthogonal polynomials with weight function $(1-s)s^{2\sqrt{\varepsilon-\beta}}$ in the closed interval $[0,1]$, giving [constant]$P_n^{(2\sqrt{\varepsilon-\beta},\,1)}(1-2s)$ [16]. By substituting $\pi(s)$ and $\sigma(s)$ in the expression $\phi(s)'/\phi(s) = \pi(s)/\sigma(s)$ and then solving the resulting differential equation, the other part of the wave function in Eq.(2) is found as

$$\phi(s) = (1-s)s^{\sqrt{\varepsilon-\beta}}, \qquad (26)$$

with $q=1$. Combining the Jacobi polynomials and $\phi(s)$ in Eq.(2), the s–state wave functions are found to be

$$R_n(s) = C_n(1-s)s^{\sqrt{\varepsilon-\beta}}P_n^{(2\sqrt{\varepsilon-\beta},\,1)}(1-2s), \qquad (27)$$

where $C_n$ is a new normalization constant determined by $\int_{-\infty}^{\infty}R_n^2(s)ds = 1$. For $n=0,1,2$ the unnormalized wavefunctions in terms of hypergeometric polynomials are shown in Fig.3.

## 3.1. PT symmetric and non-Hermitian Woods-Saxon case

We are now going to consider different forms of the standard Woods–Saxon potential, namely at least one of the parameters is imaginary. For a special case, we take the potential parameters in Eq.(10) as $V_0 \to V_0$ and $\alpha \to i\alpha_I$. Such a potential is called as $PT$–Symmetric but non–Hermitian and its shape becomes

$$V(r) = -V_0\left(\frac{1+q\cos(2\alpha_I r)-iq\sin(2\alpha_I r)}{1+q^2+2q\cos(2\alpha_I r)}\right). \qquad (28)$$

By substituting this potential into Eq.(11) and making similar operations in obtaining Eq.(23), we get the energy eigenvalues

$$E_n = \frac{2\hbar^2}{m}\left[\frac{mV_0}{4\hbar^2\alpha_I(n+1)} - \frac{\alpha_I(n+1)}{2}\right]^2, \qquad (28)$$

from the comparison of $\lambda$ values of

$$\lambda = -q\left(1+\beta+2i\sqrt{\varepsilon-\beta}\right), \qquad (30)$$

$$\lambda = \lambda_n = nq\left(3+2i\sqrt{\varepsilon-\beta}\right) + n(n-1)q. \qquad (31)$$

A positive energy spectra is obtained if and only if $n < \sqrt{\frac{mV_0}{2\alpha_I^2}}-1$, since the energy eigenvalues of Woods–Saxon potential are negative. By choosing parameter $\alpha$ as purely imaginary, the energy eigenvalues obtained for PT–symmetric and non–Hermitian Woods-Saxon potential are not similar to Eq.(24).

## 3.2. Non-PT symmetric and non-Hermitian Woods-Saxon case

Another form of the potential is obtained by making the selections of $V_0 \to iV_{0I}$ and $\alpha \to i\alpha_I$. It takes the form

$$V(r) = -V_0 \left( \frac{q\sin(2\alpha_I r) + i(1 + q\cos(2\alpha_I r))}{1 + q^2 + 2q\cos(2\alpha_I r)} \right) \qquad (32)$$

and called as non−$PT$ symmetric and also non−Hermitian. According to Refs. [17,18], this type of the complex potential is a pseudo−Hermitian and its basic properties are studied intensively.

Substituting Eq.(33) into Eq.(11), we get the value of the $\pi(s)$ as

$$\pi(s) = -\frac{qs}{2} \pm \frac{i}{2} \begin{cases} (2\sqrt{\varepsilon - i\beta} + i)qs - 2\sqrt{\varepsilon - i\beta} & \text{for } k = i\beta q + iq\sqrt{\varepsilon - i\beta}, \\ (2\sqrt{\varepsilon - i\beta} - i)qs - 2\sqrt{\varepsilon - i\beta} & \text{for } k = i\beta q - iq\sqrt{\varepsilon - i\beta} \end{cases} \qquad (33)$$

and after choosing the appropriate $k$ and $\pi$, we can write $\tau$ as

$$\tau(s) = (1 + 2i\sqrt{\varepsilon - i\beta}) - qs(3 + 2i\sqrt{\varepsilon - i\beta}). \qquad (34)$$

Thus, the energy eigenvalues are reduced to the following form

$$E_n = -\frac{2\hbar^2}{m}\left[\left(\frac{mV_0}{4\hbar^2 \alpha_I (n+1)}\right)^2 - \left(\frac{\alpha_I (n+1)}{2}\right)^2 + i\frac{mV_0}{4\hbar^2 \alpha_I}\right], \qquad (10)$$

but it has real plus imaginary energy spectra. Here, $\alpha_I$, is an arbitrary real parameter and $i = \sqrt{-1}$. When we consider the real part of energy eigenvalues an acceptable result is obtained when $n < \sqrt{\frac{mV_0}{2\alpha_I^2}} - 1$ condition. However, the energy spectrum is not seen at the imaginary part of energy eigenvalues, since it is independent of $n$.

## 4. Conclusions

The exact solution of the radial schrödinger equation with the Woods-Saxon potential for the s-states. Nikiforov−Uvarov method is used in the calculations after transforming Schrödinger equation into the hypergeometric type. Schrödinger equation is also solved for the complex potential case. The energy eigenvalues obtained for real case are compared with the ones for the complex case. According to the complex quantum mechanics [19], the eigenvalues of the conversion $\alpha \to i\alpha_I$ are not simultaneously eigenstates of $PT$ operator. If one lets $\alpha \to i\alpha_I$ in the Woods−Saxon potential, it is found that the energy levels of the potential are positive in contrary to expectation.

When $\alpha$ and $V_0$ parameters are purely complex, it is seen that the number of discrete levels for bound states is given only by the real part of energy eigenvalues. Therefore, if all the parameters of potential remain purely real, it is clear that all bound energies $E_n$ with $n \geq 0$ represent a bound state energy spectrum.

# Figure Captions

**Figure 1:** Variation of the Woods-Saxon potential as a function of r. The parameters take values $V_0 = 5$ MeV, $R_0 = 5.8$ fm and $a = 0.68\,fm$. $q$ is an arbitrary parameter.

**Figure 2:** The variation of the energy eigenvalues with respect to the quantum number $n$ with $V_0 = 5$ MeV. The curves correspond to the different values of the range parameter $a$.

**Figure 3:** Variation of the unnormalized wave functions against the exponential range parameter $s$ for the first three s-states defined in terms of hypergeometric polynomials.

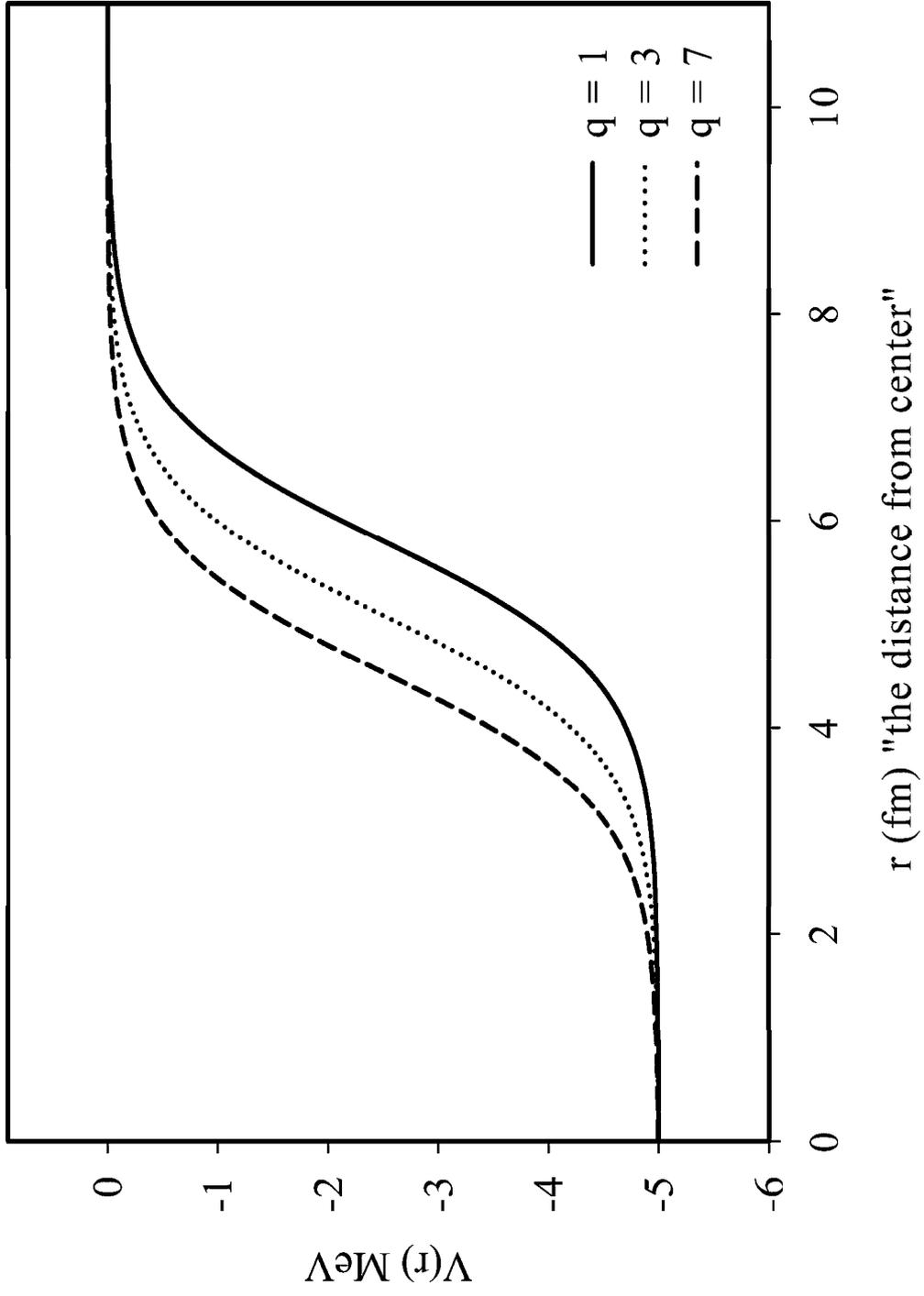

Figure 1

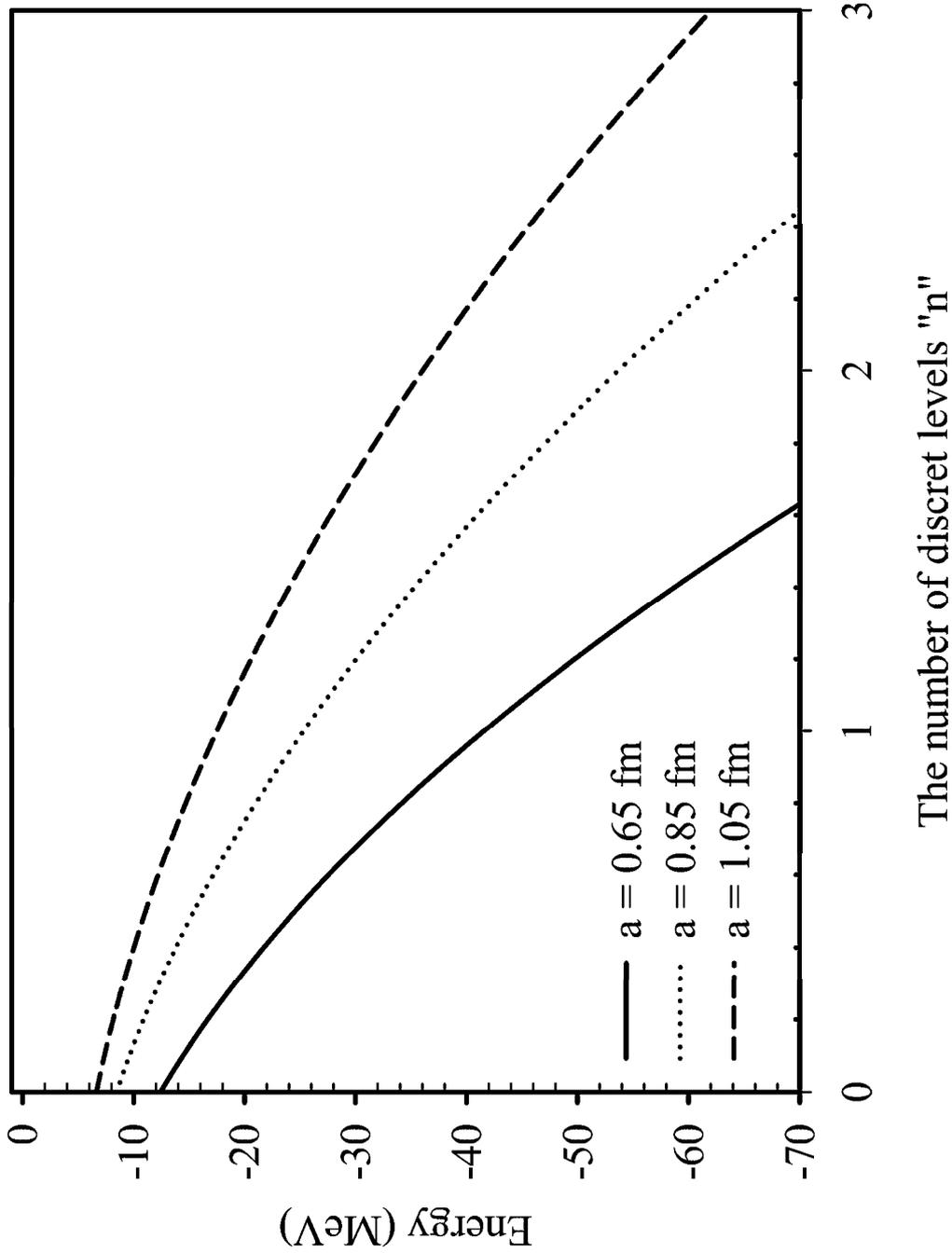

Figure 2.

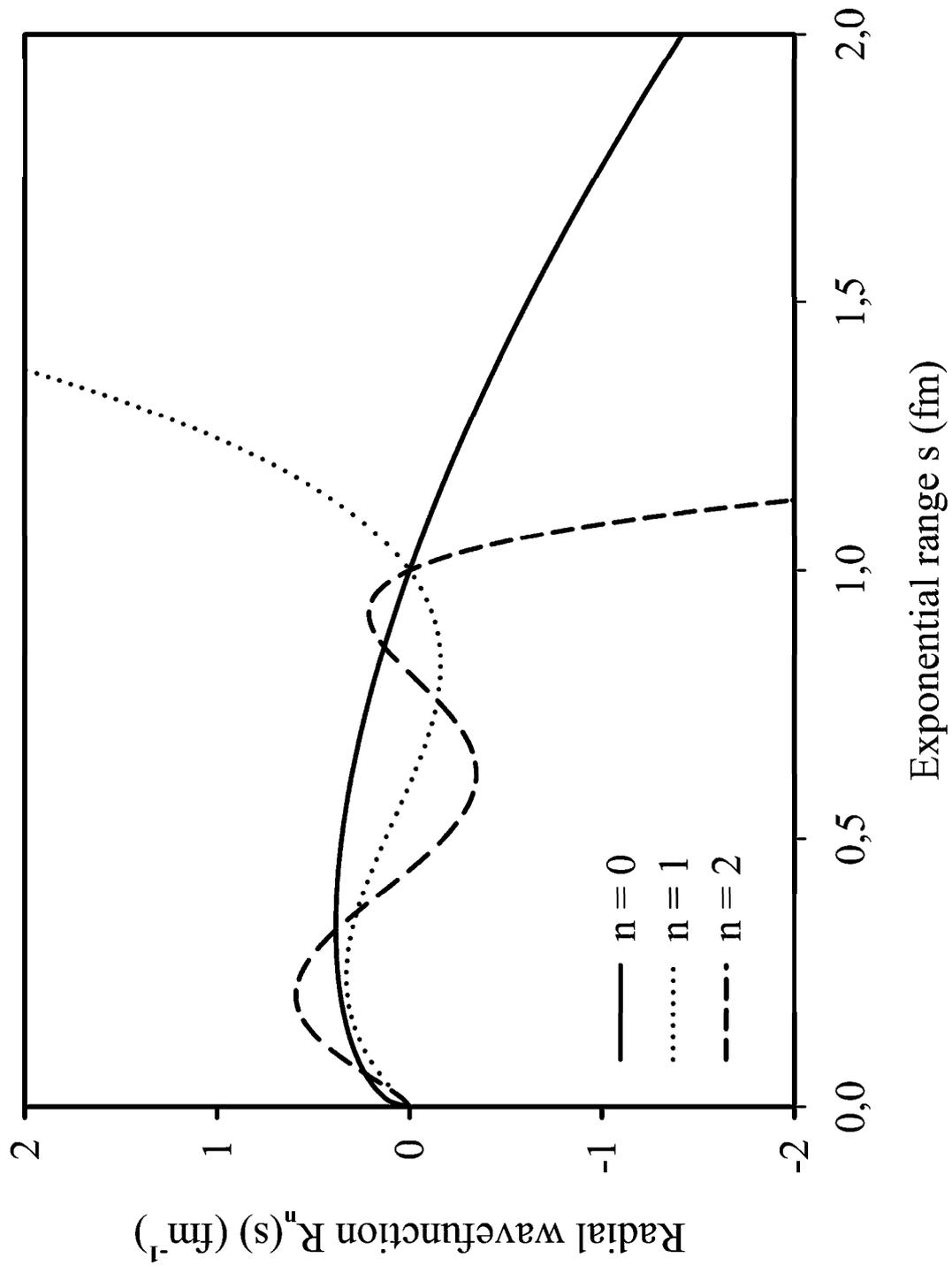

Figure 3.